\documentclass{elsart5p}

\usepackage{graphics}
\usepackage{graphicx}
\usepackage{amssymb}

\newcommand{\bra}{\langle}
\newcommand{\ket}{\rangle}
\newcommand{\cl}{\chi_{\mathrm{loc}}}

\newcommand{\mloc}{m_{\mathrm{loc}}}
\newcommand{\ra}{\rightarrow}
\newcommand{\w}{\omega}
\DeclareMathAlphabet{\bi}{OML}{cmm}{b}{it}
\newcommand{\kk}{\bi{k}}
\newcommand{\qq}{\bi{q}}

\begin{document}
\begin{frontmatter}

\title{Kondo physics in a dissipative environment}

%

\author[AA]{M.~T.~Glossop\corauthref{Glossop}},
\ead{glossop@phys.ufl.edu}
\author[BB]{N.~Khoshkhou},
\author[AA]{K.~Ingersent}

\address[AA]{Department of Physics, University of Florida, Gainesville,
Florida 32611-8440, USA}
\address[BB]{Department of Physics, Wesleyan University, Wesleyan Station,
Middletown, Connecticut 06459, USA }

\corauth[Glossop]{Corresponding author. Fax (352) 846-0295}

\begin{abstract} 
We report nonperturbative results for the interacting quantum-critical behavior
in a Bose-Fermi Kondo model describing a spin-$\frac{1}{2}$ coupled both to a
fermionic band with a pseudogap density of states and to a dissipative bosonic bath.
The model serves as a paradigm for studying the interplay between Kondo physics and
low-energy dissipative modes in strongly correlated systems.

\end{abstract}


\end{frontmatter}


\section{Introduction}
\label{Introduction}

Impurity quantum phase transitions (QPTs) and associated local non-Fermi
liquid behavior \cite{Vojta:06} presently receive much attention in
connection with tunable quantum criticality in mesoscopic devices, the physics
of nonmagnetic impurities in cuprate superconductors, and (via dynamical
mean-field theory and its extensions \cite{Si:01}) heavy-fermion quantum
criticality.
Impurity QPTs have been studied in several models involving fermionic and/or
bosonic baths. In the pseudogap Kondo model \cite{Withoff:90}, for example, a
depletion of the electronic density of states around the Fermi level can
destroy the Kondo effect that is ubiquitous for metallic hosts.
In the Bose-Fermi Kondo model (BFKM), the Kondo effect is destroyed by a
competing coupling of the impurity spin to a dissipative bosonic bath
representing collective excitations of the environment
\cite{Zhu:02,Glossop:05+07}.

In this work, we study band depletion and dissipation effects together in a
\textit{pseudogap} BFKM. The Ising-symmetry BFKM Hamiltonian is
\begin{eqnarray}
\label{H_BFK}
\hat{H}=\sum_{\kk,\sigma}\epsilon_{\kk}c^{\dagger}_{\kk\sigma}c_{\kk\sigma}
&+&\mbox{$\frac{1}{2}$}J_0\bi{S} \cdot \hspace{-0.25cm}
\sum_{\kk, \kk', \sigma, \sigma'} c_{\kk\sigma}^{\dagger}
\mbox{\boldmath$\sigma$}_{\sigma\sigma'}c_{\kk'\sigma'}\\
&+&\sum_{\qq}\w_{\qq}\phi^{\dagger}_{\qq}\phi_{\qq}+g_0 S_z
\sum_{\qq}(\phi_{\qq} + \phi^{\dagger}_{-\qq}).\nonumber
\end{eqnarray}
$J_0$ is the local Kondo exchange coupling between a local spin-$\frac{1}{2}$
$\bi{S}$ and the fermionic band, while the dissipation strength $g_0$ couples
$S_z$ to a bath of bosonic oscillators characterized by a power-law density of
states
\begin{equation}
\eta(\w)=\sum_{\qq}\delta(\w-\w_{\qq})
    =\frac{K_0^2}{\pi}\left(\frac{\w}{\w_0}\right)^s \Theta(\w)\Theta(\w_0-\w).
\end{equation}
In the pseudogap BFKM, the fermionic density
of states has a power-law pseudogap at the Fermi level ($\epsilon=0$):
\begin{equation}
\label{fermionic_DOS}
\rho(\epsilon)=\sum_{\kk}\delta(\epsilon-\epsilon_{\kk})
        =\rho_0\left|\frac{\epsilon}{D}\right|^r\Theta(D-|\epsilon|).
\end{equation}
For convenience, we set $D=\w_0=1$, in which case the model is fully
specified by the exponents $r$ and $s$ and by the dimensionless
couplings $J\equiv\rho_0J_0\ge 0$ and $g\equiv|K_0g_0|$.

The pseudogap BFKM with isotropic couplings to the bosonic bath has
been studied via perturbative RG methods \cite{Kircan:03+04}.
Such methods break down for Ising bosonic couplings, where (for $r=0$ at
least), the critical physics occurs at large $J$ and $g$.
We therefore treat Eq.\ (\ref{H_BFK}) nonperturbatively using a
recent extension of the numerical RG \cite{Glossop:05+07}.

\section{Results}

\begin{figure}
\centerline{\includegraphics[width=5cm]{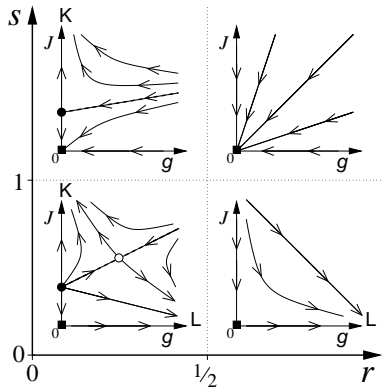}}
\caption{\label{fig1}%
Schematic RG flows in four regions of the plane spanned by the band
exponent $r$ and the bath exponent $s$. See text for discussion.}
\end{figure}

\textbf{RG flows:}
Figure \ref{fig1} shows the qualitative dependence of the RG flows on the
exponents $r\ge 0$ and $s>0$.
With $g=0$, Eqs.\ (\ref{H_BFK}) and (\ref{fermionic_DOS}) describe the
pseudogap Kondo model.
For $0<r<\frac{1}{2}$, the stable Kondo fixed point (K) is
reached only for $J>J_c(g=0)$.
For $J<J_c(0)$, flow is towards the free-impurity fixed point (FI, shown as a
square in Fig.\ \ref{fig1}) at which the impurity decouples from the baths.
The transition between the K and FI phases occurs at the fermionic critical
point (FC, solid circle in Fig.\ \ref{fig1}), with properties that are well
understood.
For $r\ra 0^+$, FC merges with FI and the impurity spin is Kondo screened for
all $J>0$.
By contrast, FC merges with K as $r\ra\frac{1}{2}^-$, and Kondo physics
is inaccessible for $r\ge\frac{1}{2}$ \cite{GBI:98}.

For $g>0$ and $s<1$, the RG flow for small $J$ is instead towards the localized
fixed point (L), at which the impurity dynamics are controlled by the dissipative
bath.
For $s<1$ and $0\le r<\frac{1}{2}$, a continuous QPT between the K and L takes
place at a second critical point (BFC, open circle in Fig.\ \ref{fig1}) lying on
the separatrix $J_c(g)$ (dashed line in Fig.\ \ref{fig1}).

The effects of dissipation lessen with increasing $s$: as $s\ra 1^-$, BFC merges
with FC, and for $s>1$ the essential physics is that of the pseudogap
Kondo model.
Increasing $r$ inhibits Kondo screening: as $r\ra \frac{1}{2}^-$, BFC merges
with K; for $r>\frac{1}{2}$ and $s<1$, the RG flow is towards L for any
nonzero $g$.

\textbf{Local magnetic response:}
The critical properties of the pseudogap BFKM reveal themselves most
clearly in the response to a local magnetic field $h$ that
acts only on the impurity spin through an additional Hamiltonian term $h S_z$.
The critical behavior at FC has been reported in \cite{Ingersent:02}.

Near BFC, the imaginary part of the local dynamical
susceptibility obeys the scaling form characteristic of an interacting
critical fixed point:
\begin{equation}
\cl''(\w,T) =
T^{\eta-1} \, \Phi( \w/T, \, j/T^{1/\nu} ) ,
\end{equation}
where $j=J/J_c-1$ measures the distance to criticality, and the exponent
$\nu$ governs the vanishing of a crossover scale $T_*\propto |j|^{\nu}$
above which quantum-critical behavior is observed up to nonuniversal energy scales.
We find that the anomalous exponent characterizing critical local-moment
fluctuations is $\eta=1-s$ independent of $r$, whereas $\nu$ exhibits both
$r$ and $s$ dependence.

Knowledge of $\eta$ and $\nu$ is sufficient to determine all critical
exponents associated with the response to $h$. Such hyperscaling behavior
is expected at an interacting quantum-critical point having an impurity
free energy of the form $F_{\mathrm{imp}}=Tf(j/T^{1/\nu},|h|/T^b)$.
For instance, the local magnetization $\mloc=\bra S_z (T=0, h \ra 0)\ket$
serves as the order parameter for the transition. It obeys
$\mloc(j<0)\propto(-j)^{\beta}$,
where $2\beta=\nu\eta$ via hyperscaling. Figure 2a shows $\mloc$ versus $j$
for fixed band exponent $r=0.2$ and three values of the bath exponent $0<s<1$;
in all cases $\mloc(j)$ vanishes continuously as $j\ra 0^-$, and $\mloc(j)=0$
for $j\ge 0$. It is clear from Fig.\ 2a, and from the logarithmic plots in
Fig 2b, that the critical exponent $\beta$ exhibits $r$ and $s$ dependence.

\begin{figure}
\centerline{\includegraphics[angle=-90,width=7.5cm]{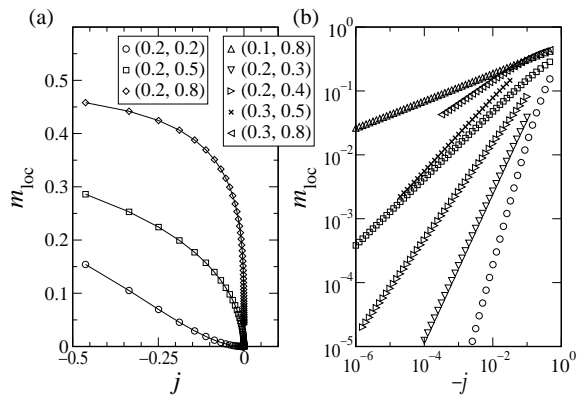}}
\caption{\label{fig2}%
Order parameter $\mloc(j)$ (defined in text) versus $j=J/J_c-1$ for
the $(r,s)$ pairs identified in the legend.}
\end{figure}

Throughout the domain $0<r<\frac{1}{2}$, $0<s<1$, the BFC exponents satisfy
hyperscaling relations to within our estimated numerical uncertainty.
For example, for $r=0.2$ and $s=0.2$, $\eta=0.8003(5)$, $1/\nu=0.200(1)$, and
$\beta=2.001(2)$, with parentheses enclosing the uncertainty in the last digit.

\section{Summary}
\label{summary}

We have studied nonperturbatively the critical properties of the impurity
quantum phase transition between Kondo-screened and localized-moment phases in
the pseudogap Bose-Fermi Kondo model. Critical exponents depend only on the
exponents parameterizing the densities of states of the fermionic band and
the bosonic bath, and are found to obey hyperscaling relations characteristic
of an interacting quantum-critical point.
Further details will be discussed in a forthcoming publication.

This work was supported in part by NSF Grants DMR-0552726 (the Univ.\ of Florida
Physics REU Site) and DMR-0312939.
Resources and support were provided by the
Univ.\ of Florida High-Performance Computing Center.

\end{document}